\documentclass[aps, prl, amsmath, floats,floatfix, twocolumn, superscriptaddress]{revtex4}

 \usepackage{xcolor}
\usepackage{amssymb}
\usepackage{amsmath}
\usepackage{verbatim}
\usepackage{mathrsfs}
\usepackage{amsfonts}
\usepackage{latexsym}
\usepackage{epsfig}
\usepackage{color}
\usepackage{graphicx,subfigure}
\usepackage{units}
\usepackage{slashbox}
\usepackage[spanish,english]{babel}
\usepackage{soul}

\usepackage{mathtools,slashed}

\newcommand{\be}{\begin{equation}}
\newcommand{\ee}{\end{equation}}
\newcommand{\bea}{\begin{eqnarray}}
\newcommand{\eea}{\end{eqnarray}}

\begin{document}

\title{{\bf Spontaneous creation of circularly polarized photons in chiral astrophysical systems}}

\author{Adrian del Rio}
\author{Nicolas Sanchis-Gual}
\affiliation{Centro de Astrof\'\i sica e Gravita\c c\~ao - CENTRA,  
Instituto Superior T\'ecnico - IST, Universidade de Lisboa - UL, Avenida
Rovisco Pais 1, 1049-001, Portugal}

 \author{Vassilios Mewes}
\affiliation{ National Center for Computational Sciences, Oak Ridge National Laboratory, P.O. Box 2008, Oak Ridge, TN 37831-6164, USA}

\affiliation{Physics Division, Oak Ridge National Laboratory, P.O. Box 2008, Oak Ridge, TN 37831-6354, USA}
\affiliation{Center for Computational Relativity and Gravitation,  School of Mathematical Sciences, Rochester Institute of Technology, 85 Lomb Memorial Drive, Rochester, New York 14623, USA}

    \author{Ivan Agullo}
\affiliation{Department of Physics and Astronomy, Louisiana State University, Baton Rouge, LA 70803-4001}

\author{Jos\'e~A. Font}
\affiliation{Departamento de
  Astronom\'{\i}a y Astrof\'{\i}sica, Universitat de Val\`encia,
  Dr. Moliner 50, 46100, Burjassot (Val\`encia), Spain}
\affiliation{Observatori Astron\`omic, Universitat de Val\`encia,  Catedr\'atico
  Jos\'e Beltr\'an 2, 46980, Paterna (Val\`encia), Spain}

 \author{Jose Navarro-Salas
}
\affiliation{Departamento de F\'isica Te\'orica and IFIC,  Universitat de Val\`encia-CSIC.  Dr. Moliner 50, 46100, Burjassot (Val\`encia), Spain.}

\date{\today}

\begin{abstract}
This work establishes a relation between chiral anomalies in curved spacetimes and the radiative content of the gravitational field. In particular, we show that a flux of circularly polarized gravitational waves triggers the spontaneous creation of photons with net circular polarization from the quantum vacuum.
 Using  waveform catalogues we identify  precessing binary black holes  as  astrophysical  configurations  that emit such  gravitational radiation, and then
solve the fully non-linear Einstein's equations with numerical relativity to evaluate the net effect. The quantum amplitude for a merger is comparable to the Hawking emission rate of the final black hole, and small to be directly observed. 
However,  the implications  for the inspiral of  binary neutron stars could be more prominent, as argued on symmetry grounds.

\end{abstract}

\maketitle

{\bf {\em Introduction.}} Maxwell equations without  charges and currents
are invariant under duality rotations of the electromagnetic field, $F_{ab}\to F_{ab}\cos\theta +{^{\star}F}_{ab} \sin\theta$.   This continuous transformation was shown to be a physical symmetry of the source-free Maxwell action, $S=-\frac{1}{4}\int_{M} d^4x \sqrt{-g} F_{ab}F^{ab}(x) $,  in a general curved spacetime background $(M, g_{ab})$  \cite{deser-teitelboim}. 
By Noether's Theorem  this symmetry leads to a  conserved  current, $\nabla_a j_D^a=0$. Its associated  Noether charge $Q_D$  
physically accounts  for the difference between right- and left-handed circularly polarized  components \cite{Calkin} ---the usual V-Stokes parameter--- as one could expect, since duality rotations are just chiral transformations for the electromagnetic field. Thus, this symmetry guarantees that the circular polarization state of classical electromagnetic waves  is a constant of motion, even if the field propagates in an arbitrary gravitational field.

However, the quantum theory offers a more interesting scenario~\cite{papers}. In presence of
gravity, quantum fluctuations of the electromagnetic field can spoil the classical conservation law as
  $\left<0|\nabla_a j_D^a|0\right>= \frac{-1}{96\pi^2}R_{abcd}{^{\star}}R^{abcd}\neq 0$, where $\left|0\right>$ is any vacuum state, and $R_{abc}^{\hspace{0.45cm}d}$  is the Riemann tensor of the spacetime. 
This is understood as the spin-1 generalization of the Fermionic chiral anomaly \cite{ABJ}.
{The goal of this paper is to explore the physical consequences of this non-conservation law, and to identify astrophysical systems where the new effect could be relevant.}

{\bf {\em Chiral flux of photons and the importance of gravitational dynamics. }}
{The anomalous current indicates that} the  Noether charge is no longer conserved in time and that its change is determined by the geometry of the background. More precisely, for asymptotically Minkowski spacetimes \cite{Ashtekar14}, where we can compare the values of the Noether charge {at} past and future null infinities, $\mathcal J^{\pm}$ (where the gravitational interaction vanishes and preferred notions of quantum vacua exist) one finds  
\begin{eqnarray}
\left<{\rm 0}|Q_D(\mathcal J^+)  |{\rm 0}\right> & - &  \left<{\rm 0}| Q_D(\mathcal J^-) |{\rm 0}\right> \label{CPintegral} \\
& = & \frac{-1}{96\pi^2} \int_{ M} d^4x\sqrt{-g}R_{abcd}{^{\star}R}^{abcd} \,. \nonumber
\end{eqnarray}
If the electromagnetic field is initially in vacuum, $\left<{\rm 0}| Q_D(\mathcal J^-) |{\rm 0}\right>=0$, 
 a non-zero value at late times, $\left<{\rm 0}|Q_D(\mathcal J^+)  |{\rm 0}\right>\neq 0$, implies that a flux of photons with net circular polarization arrives at $\mathcal J^+$.  This happens  only if the right hand side (RHS) of (\ref{CPintegral}) is different from zero. This is  called the Chern-Pontryagin,  and it measures the degree 
of  spacetime chirality.    It is dimensionless, as the expectation value of the number operator written in the left hand side.
 The helicity of photons can be measured by direct detections at $\mathcal J^+$ through polarimetry observations, and hence a good understanding of the RHS is fundamental to make precise predictions. 

While a non-zero value of the Chern-Pontryagin is  usually associated  with the presence of certain gravitational instantons \cite{EHG}, these are however    solutions of the Euclidean Einstein's equations only, 
whose  interpretation as quantum-tunneling  between different topologies \cite{GH77} relies on  quantum gravity considerations. {Our interest is in solutions of a more direct physical interpretation}. We aim to identify   Lorentzian geometries {satisfying Einstein's equations}  having  a non-zero Chern-Pontryagin, and then to calculate (\ref{CPintegral})  for such classical gravitational fields.   
The identification of these systems is   non-trivial, since physically realistic exact  solutions to Einstein's equations 
provide a null contribution to (\ref{CPintegral}).  For instance, 
black hole (BH) solutions belonging to the Kerr-Newman family yield a vanishing integral.
  This is because the Chern-Pontryagin is a pseudoscalar (flips sign under improper rotations) while the underlying Kerr-Newman metric is even under mirror reflections with respect to the  plane perpendicular to the symmetry axis.   We must consider  spacetimes with no mirror symmetries.
 Uniqueness theorems in general relativity \cite{CCH12} assert that any regular, asymptotically flat, stationary vacuum solution in four dimensions is  a member of the above family. Since we are interested in asymptotically flat spacetimes ---to have a good notion of photons and helicity  at asymptotic times--- the above considerations suggest that relevant spacetimes will only come up  if we give up stationarity.

The lack of {stationarity}
 makes  the analysis highly non-trivial, and would require in general numerical methods. 
In the following  we  provide a physical characterization of   spacetimes having a nonzero Chern-Pontryagin,  finding in turn a deep interplay between handedness of photons and emission of circularly polarized gravitational-waves (GW). This is a new and purely gravitational effect, in sharp contrast to the instanton solutions reported in the literature (which are only topological, not geometrical).

{\bf {\em Flux of circularly polarized GW}}. 
By the Chern-Weyl Theorem from the theory of characteristic classes \cite{Nakahara}, the difference $R_{abcd}{^{\star}}R^{abcd}-R'_{abcd}{^{\star}}{R'}^{abcd}$, with $R_{abcd}$ and $R'_{abcd}$  the curvature tensors   of any two  connection 1-forms $\omega$ and $\omega'$ on $M$, is {\it exact}
\bea
d^4x\sqrt{-g}R_{abcd}{^{\star}}R^{abcd}-d^4x\sqrt{-g'}R'_{abcd}{^{\star}}{R'}^{abcd}=dT(\omega,\omega'), \nonumber
\eea
 and given by the so-called transgression term $T(\omega,\omega')$, whose explicit expression in terms of the difference $\theta:=\omega-\omega'$ can be found in \cite{EHG}:
\bea
T(\omega,\omega')={\rm Tr} (2\theta \wedge R+\frac{2}{3}\theta\wedge \theta\wedge \theta-2\theta \wedge\omega \wedge \theta-\theta \wedge d\theta)\, .\nonumber
\eea
Taking $\omega'$ as the flat connection of Minkowski spacetime \cite{footnote-2.5}, we have $R'_{abcd}{^{\star}}{R'}^{abcd}=0$ 
\cite{footnote-3}.
Consider now $M_s=\{p\in M/ r(p)\leq s\}\subset M$, with $r$  a radial coordinate centered at the sources, and $s$  a sufficiently large number so that all gravitational sources are well inside $M_s$. We can  use  Stokes theorem  to write the volume integral of $M_s$ as the integral over the boundary,
\bea
\int_{ M_s} d^4x\sqrt{-g}R_{abcd}{^{\star}R}^{abcd} =\int_{\partial M_s}  \label{aux} T(\omega,\omega')\,.
\eea
 We now take the limit $r\to \infty$. If we assume no incoming GWs, and that all gravitational sources have spatial compact support (corresponding to isolated bodies in astrophysics) then we expect  non-trivial contributions only from future null infinity.  Following Bondi-Sachs \cite{BMS62} we  foliate  the  asymptotic  region  of $M$ around $\mathcal J^+$ by  outgoing  null  hypersurfaces $u=$  const (physically representing retarded time) each one generated by null geodesics of affine parameter $r$, and  take the limit $r\to \infty$ along these geodesics to reach $\mathcal J^+$.
Using then the Newman-Penrose spin-coefficient formalism \cite{NP62} and the asymptotic properties developed in \cite{NU68}, one can  find that \cite{followup}
\bea
\frac{-1}{96\pi^2}\int_{ M} d^4x\sqrt{-g}R_{abcd}{{^\star R}}^{abcd}  = \int_{ -\infty}^{\infty}du f(u)  \equiv \Delta Q_{\mathcal J^+\,,} \label{eb_eq}
\eea
where \cite{footnote-4}
\begin{eqnarray}
f(u):=\frac{1}{72\pi^2}\int^{u}_{-\infty} du' \times \label{ebu} && \\
 \sum_{\ell m} [ { \rm Re} \Psi_{4,\ell m}^0(u) { \rm Im} \Psi_{4,\ell m}^0(u') &- &{ \rm Re} \Psi_{4, \ell m}^0(u') { \rm Im} \Psi_{4, \ell m}^0(u) ].  \nonumber
\end{eqnarray}
In this formula $\Psi_4^0(u,\theta,\phi)=\lim_{r\to \infty} r \Psi_4(u,r,\theta,\phi)$ is the leading order coefficient of the Weyl scalar  $\Psi_4$ \cite{NP62}, and we have expanded the angular part in spherical harmonics of modes $(\ell,m)$ and spin weight $-2$. 
{Although this expression may look complicated, its physical interpretation is remarkably simple.}
Recall that the usual GW polarization modes $h_+$, $h_\times$  are in correspondence with  $\Psi_4^0$ by $\partial_u^2 h_+={\rm Re}\Psi_4^0$, $\partial_u^2 h_\times={\rm Im}\Psi_4^0$. If we expand  $h_{\times}$ and $h_+$ in Fourier modes,  then (\ref{eb_eq}) is equal to
\bea
 \int_0^{\infty} \frac{d\omega \omega^2}{72\pi^2} \sum_{\ell m} \left[| h_+^{\ell m}(\omega)+ih^{\ell m}_{\times}(\omega)|^2-|h_+^{\ell m}(\omega) - ih^{\ell m}_{\times}(\omega)|^2 \right]\, , \nonumber
 \eea
which measures the
difference  in the intensity between  right and left  circularly-polarized GW reaching future null infinity. 
{Thus,  from (\ref{CPintegral}) we conclude that the emission of chiral gravitational radiation by astrophysical systems implies  the spontaneous creation of photons with net chirality.} The more right(left)-handed GWs the spacetime contains, the more left(right)-handed photons will be excited from the quantum vacuum. 

{In the presence of a BH, $\mathcal J^+$ alone is not a Cauchy hypersurface, and one needs to add the contribution from the future event horizon $H$ to (\ref{eb_eq})-(\ref{ebu}), i.e. of the information that does not escape to infinity but rather falls into the singularity. The contribution of $H$ can be computed in a similar manner but the final expression is complicated and does not have a simple interpretation. However, it can be checked that for an isolated horizon \cite{ABL02} the contribution is zero, and from this one infers that a flux of (chiral) gravitational perturbations across $H$ is needed in order to contribute to the Chern-Pontryagin.}

{\bf {\em Identification of chiral astrophysical systems using GW data.}} 
A non-vanishing value of (\ref{ebu}) can be used as a sufficient condition to identify chiral systems. 
Mergers of binary black holes (BBH) and binary neutron stars (BNS), as those detected by advanced LIGO and Virgo~\cite{GWTC-1}, are promising candidates. 
Since numerical-relativity  simulations of compact binary mergers are computationally expensive, we first compute Eqs.~(\ref{eb_eq}) and (\ref{ebu}) using GW data from freely available catalogues~\cite{Lovelace,Healy7,Healy8,Dietrich1} to identify suitable spin and mass configurations. BBH catalogues are extensive and this
  allows us to efficiently cover a large  range of parameter space; BNS simulations however are scarce and only some conclusions can be drawn at present. With the most suitable BBH candidates, we perform numerical simulations in the following section to calculate the  integral  (\ref{CPintegral}).

\begin{table}
\begin{center}
\scalebox{0.91}{
\begin{tabular}{|c|c|c|ccc|ccc|c|}
\hline
Model & & $q$  & $\chi_{1x}$ & $\chi_{1y}$ & $\chi_{1z}$ & $\chi_{2x}$ & $\chi_{2y}$ & $\chi_{2z}$ & $72 \pi^2 \Delta Q_{\mathcal J^+}$ \\
\hline
BBH-0001 & NS & 1.00 & 0 &0 &0 &0 &0 &0                                            &-4.0$\times10^{-14}$ \\
BBH-0038 & P & 1.00 & 0.77	&0&	-0.20&	0&	0&	0.20             & 1.7$\times10^{-1}$\\
BBH-0062 & A & 0.82	  &0	&0	&-0.44	&0	&0	   &   0.33	                  & 9.4$\times 10^{-13}$\\
BBH-0068& A &1.00& 0 &0 &-0.80 &0 &0 &-0.80                                               &  -6.5$\times10^{-13}$\\
BBH-0105& A & 0.33  &0	&0	&-0.80	&0	&0	&0	                    & 2.9$\times10^{-11}$\\
BBH-0130 & P & 0.75	 &0	&0	&-0.80	&0.53	&0	&0.60           &       1.4$\times10^{-1}$\\
BBH-0136 & P & 0.50	 &0	&0	&0	&0.42	&0	&0.42	                   &3.3$\times10^{-2}$\\
BBH-0137 & P & 0.50 &-0.35	&0	&0.35	&0.35	&0	&0.35  &  1.2$\times10^{-1}$\\
BBH-0272 & P & 0.67 &	0 &	0 &	0 & 0& 	0.69& 	0.40& -2.2$\times10^{-1}$\\
\hline
BNS(oo) & NS &0.82 &0 &0 &0 &0 &0 &0 & 6.2$\times10^{-5}$\\
BNS(uu) & A    &0.82 &0 &0 &0.08 &0 &0 &0.09 &-1.3$\times10^{-4}$\\
BNS(dd) & A    &0.82 &0 &0 &-0.08 &0 &0 &-0.09 &-5.1$\times10^{-5}$\\
BNS(nn) & P    & 0.82&0.08 &0.08 &0.08 &0.09&0.09 &0.09 & 2.0$\times10^{-4}$\\
BNS(ss) & P    & 0.82&-0.08 &-0.08 &-0.08 &-0.09 &-0.09 &-0.09  &-7.4$\times10^{-5}$\\
\hline
\end{tabular}
}
\end{center}
\caption{ Values of  $\Delta Q_{\mathcal J^+}$ for a representative sample of BBH (obtained from the RIT catalogue~\cite{Lovelace,Healy7,Healy8}) and some  BNS mergers (obtained in~\cite{Dietrich1}, with the SLy equation of state).  Types `NS', `A' and `P' correspond to non-spinning, aligned and precessing binaries, respectively.  The spins of the two initial objects, $\vec \chi_1$ and $\vec \chi_2$, are shown in Cartesian coordinates, and $q=m_1/m_2$ indicates the mass ratio. The GW peak luminosity for all BBH models fluctuates around $\sim$ $10^{56}$ erg s$^{-1}$.  BBH-0062  corresponds to LIGO event GW150914. } 
\label{tabbbhgw}
\end{table}

 Results from illustrative models of BBH mergers are shown in Table~\ref{tabbbhgw}.  We find no clear dependence with mass ratio, BH spins, merger time or GW luminosity.
Interestingly, only precessing systems produce  non-negligible values. These occur when the spins of the BHs are misaligned with the binary's orbital angular momentum. 
Mergers of spin-aligned objects yield  values compatible with zero because positive and negative $m$ modes are found to cancel   each other up to roundoff error for all  $\ell$  in Eq.~(\ref{ebu}). This result can  be understood as follows.
Consider the functional $F[g_{ab}]=\int_{\Sigma} d^3x\sqrt{-g}R_{abcd}{^*R}^{abcd}$. As  a pseudo-scalar it changes sign under a reflection $I$ (improper rotation) and remains invariant under a proper rotation $R$, namely $(R\circ I)(F[g_{ab}])=F[(R\circ I)g_{ab}]=-F[g_{ab}]$. Suppose now a binary system with metric $g_{ab}$, in which the two spins remain always aligned;   
by taking a mirror reflection with respect to the separation plane,
we can always return to the same physical configuration $g_{ab}$ after  a suitable rotation. 
In other words, $F[(R\circ I)g_{ab}]=F[g_{ab}]$. This  leads to $F[g_{ab}]=0$, which explains the observed result. This argument fails if the two spins are misaligned
and justifies the observed non-zero values in Table~\ref{tabbbhgw}  for precessing binaries.

\begin{figure}[t]
\centering
\includegraphics[scale=0.3,trim=0.1cm 0cm 0cm 0cm]{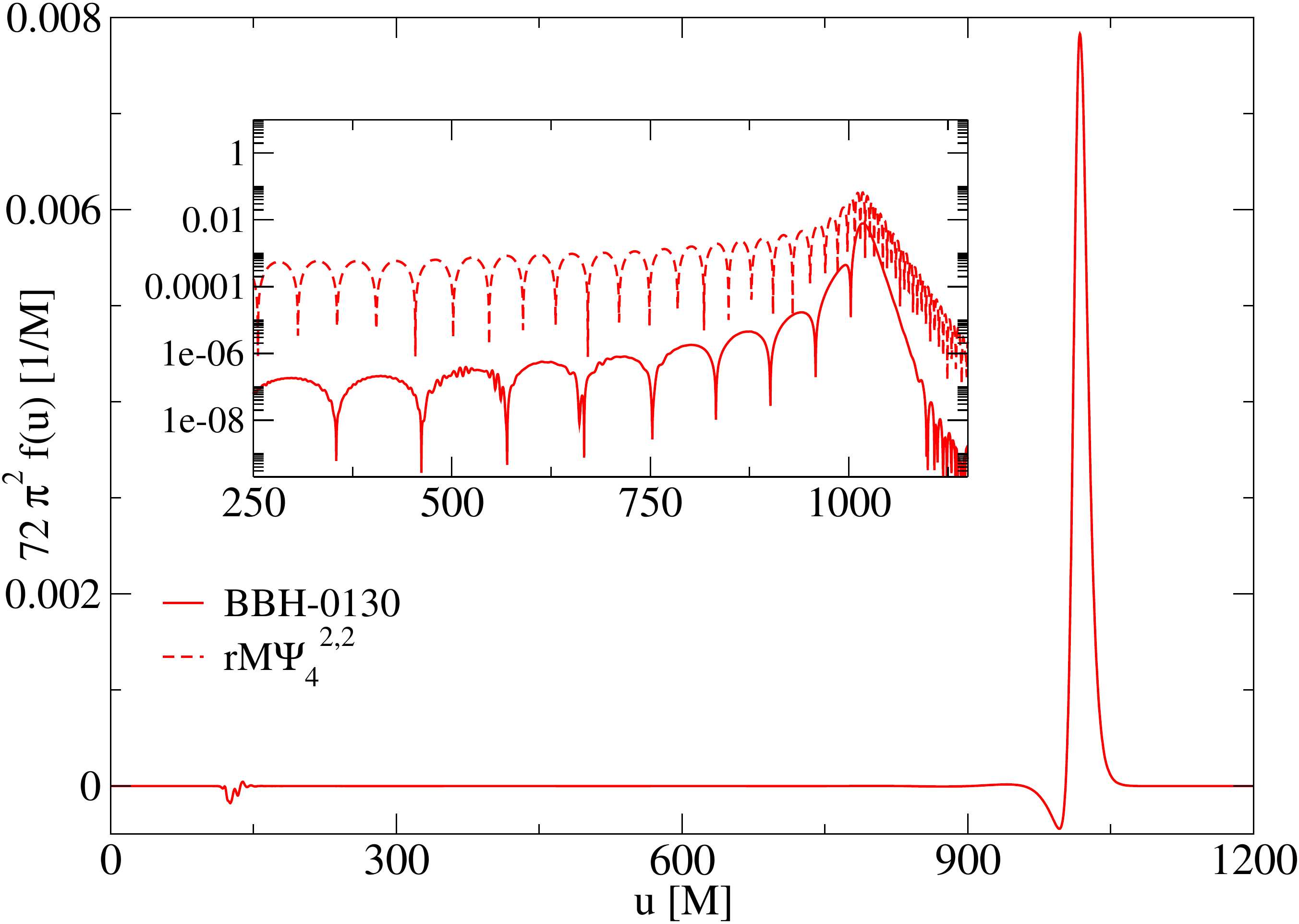}\hspace{0.5cm} 
\caption{ Amplitude, computed from Eq.~(\ref{ebu}), of GW circular polarization received at $\mathcal J^+$ as a function of retarded time $u$ for a precessing BBH waveform of Table~\ref{tabbbhgw} (model BBH-0130).  A comparison with the dominant mode $(\ell=m=2)$ of $\Psi_4$ is shown in the inset for the same model.  $M$ is the sum of the two initial horizon masses. } 
\label{fig.2}
\end{figure}

Table~\ref{tabbbhgw} also shows that the values obtained for some available BNS mergers~\cite{Dietrich1} are 3-4 orders of magnitude smaller than those for precessing BBH mergers. This is because, despite the orbital plane may precess, the individual neutron star spins are very low and  aligned (in all available simulations).  The fact that the values are not compatible with zero (unlike the BBH case) must be due to the complex multipolar structure of BNS mergers, which breaks the perfect symmetry under mirror reflections. In this respect,  notice that even a  merger of non-rotating NS leads to a nonzero value.

The value of  $\Delta Q_{\mathcal J^+}$ in Table~\ref{tabbbhgw} captures the total chirality of GW that reach  ${\mathcal J^+}$. It is also interesting to study the flux of chirality per unit of retarded time, given by (\ref{ebu}). This is displayed in Fig.~\ref{fig.2} for  BBH-0130.  The time evolution shows that the amplitude $f(u)$ oscillates around zero during the inspiral  and it peaks at the time of merger, in coincidence with the largest GW burst. The detection of this peak  would be a clear indication of the precessing character of the binary.

{\bf {\em Binary black hole simulations -- contribution of the BH horizon.}} 
We next perform numerical simulations of BBH mergers to gauge the contribution of the BH horizon to the total Chern-Pontryagin. We solve Einstein's equations numerically for the spacetime metric and calculate (\ref{CPintegral}). The 4D integral  can be easily split in the usual 3+1 formalism employed in numerical relativity~\cite{alcubierre} by noticing that $R_{abcd}{^\star R}^{abcd}=16E_{ab}B^{ab}$, where $E_{ab}$ and $B^{ab}$ are the electric and magnetic parts of the Weyl tensor, respectively. These are  purely spatial tensors, so the task reduces to calculate the integral between two spacelike Cauchy hypersurfaces $\Sigma_1$ and $\Sigma_2$:
\begin{eqnarray}\label{DQ}
\left<{\rm 0}|Q_D(\mathcal J^+)  |{\rm 0}\right>  = \frac{-1}{6\pi^2}\int_{t_1}^{t_2}dt \int_{\Sigma_t\subset \mathbb R^3}d\Sigma  \sqrt{-g}  E_{ab}B^{ab}\, .
\end{eqnarray}
We compute Eq.~(\ref{DQ}) by modifying the {\tt Antenna} thorn~\cite{Antenna1} and perform simulations using the {\tt Einstein Toolkit}~\cite{ET}, using the {\tt McLachlan} thorn~\cite{McLachlan1} for the spacetime evolution. To simulate quasi-circular BBH mergers, we take the component masses and initial linear momentum from~\cite{Tichy}, in particular those corresponding to BHs initially separated by $D/2=\lbrace4.5M,8.1M\rbrace $ in units of the ADM mass $M$. The initial data are calculated using the {\tt TwoPunctures} thorn~\cite{Ansorg}. In both cases we consider Kerr BHs with spins tilted 45$^{\circ}$ with respect to the orbital plane and 90$^{\circ}$ with respect to each other: $(\leftarrow,0,\uparrow),\, (\rightarrow,0,\uparrow)$. The results  are shown in Table~\ref{tabhdbbh}.

\begin{figure}[t!]
\centering
\includegraphics[scale=0.093,trim=0.1cm 0cm 0cm 0cm]{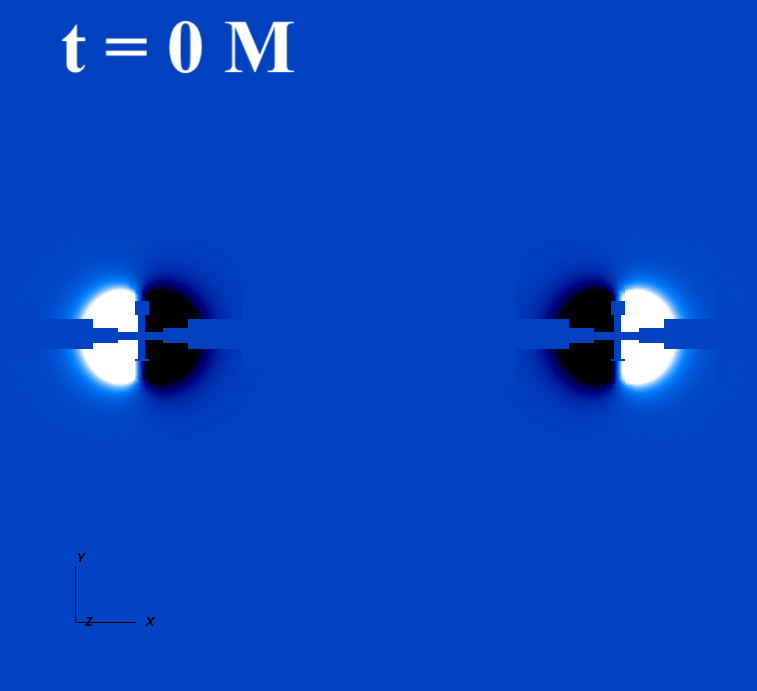} \includegraphics[scale=0.093,trim=0.1cm 0cm 0cm 0cm]{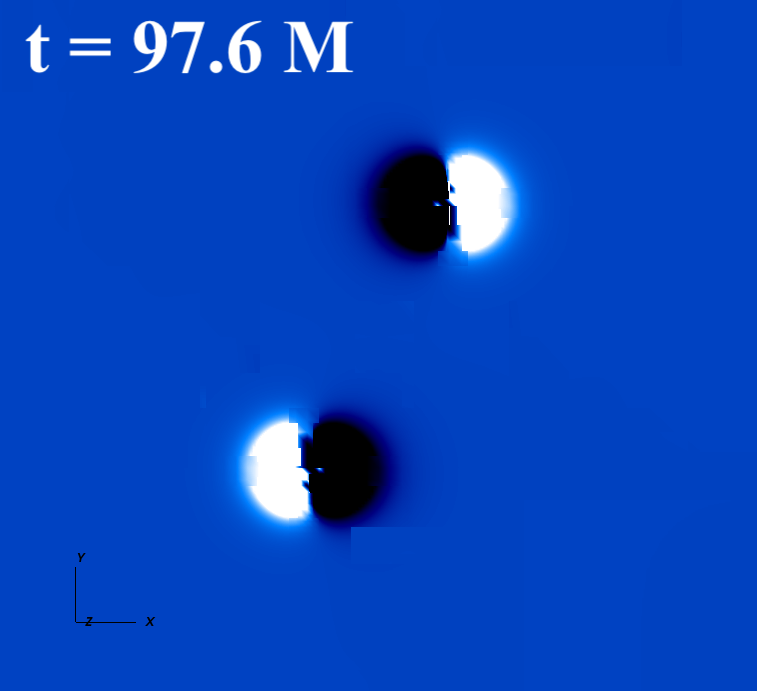}
\includegraphics[scale=0.093,trim=0.1cm 0cm 0cm 0cm]{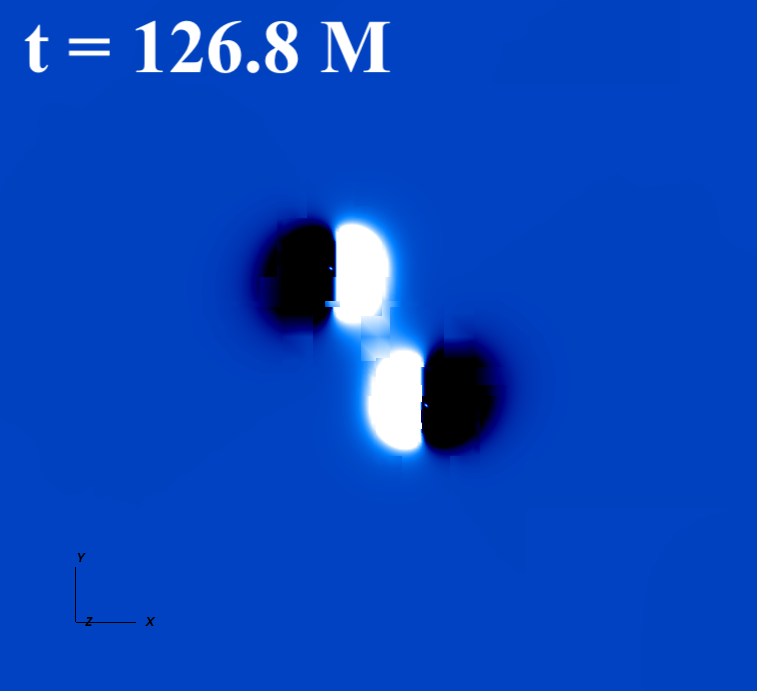}
\includegraphics[scale=0.093,trim=0.1cm 0cm 0cm 0cm]{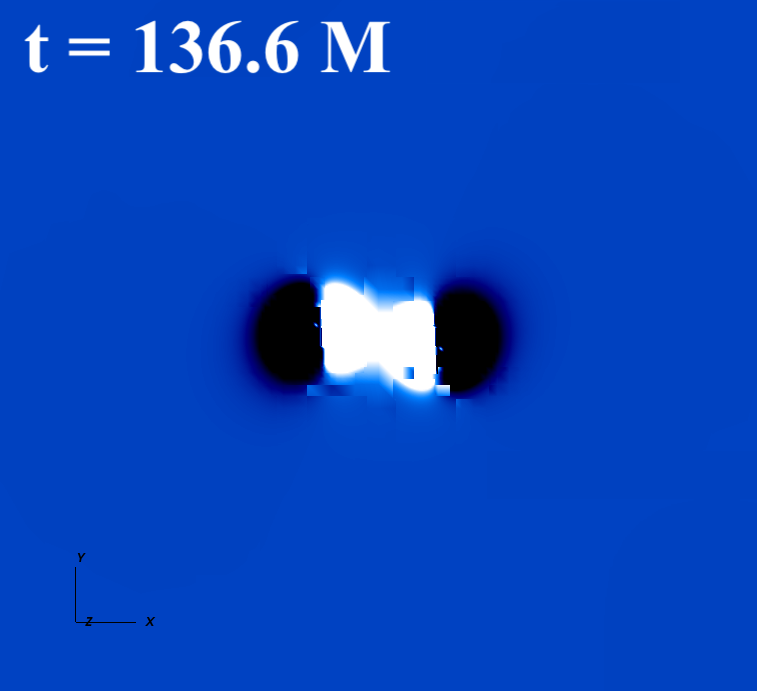}
\includegraphics[scale=0.093,trim=0.1cm 0cm 0cm 0cm]{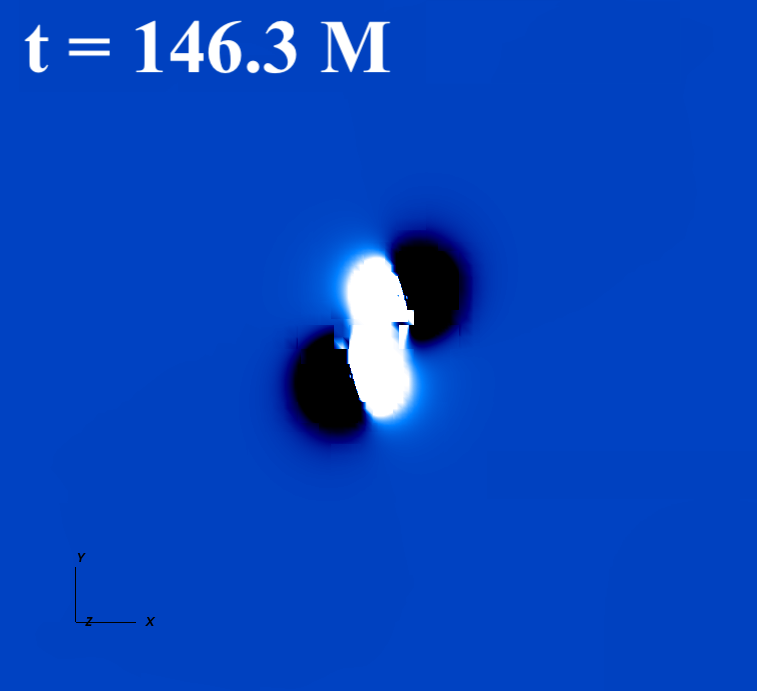}
\includegraphics[scale=0.093,trim=0.1cm 0cm 0cm 0cm]{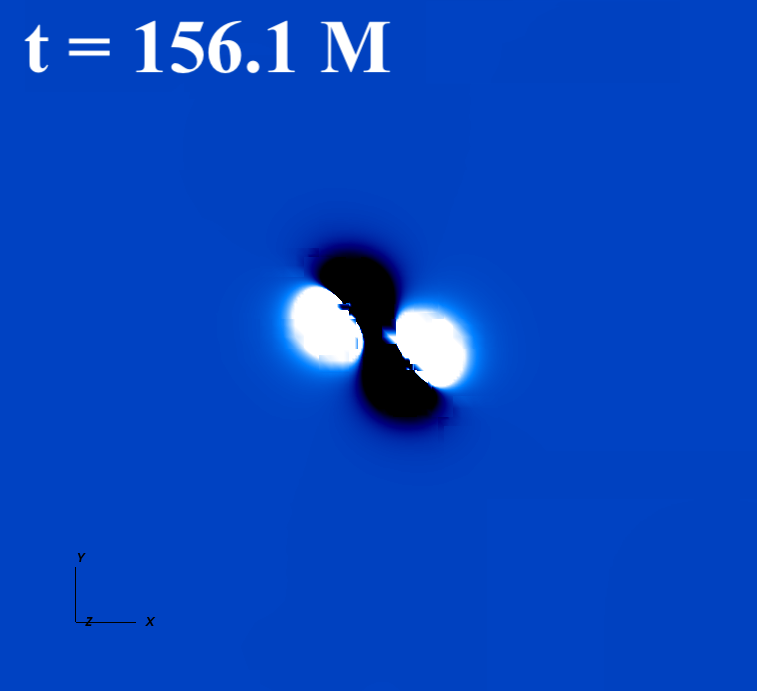}
\includegraphics[scale=0.093,trim=0.1cm 0cm 0cm 0cm]{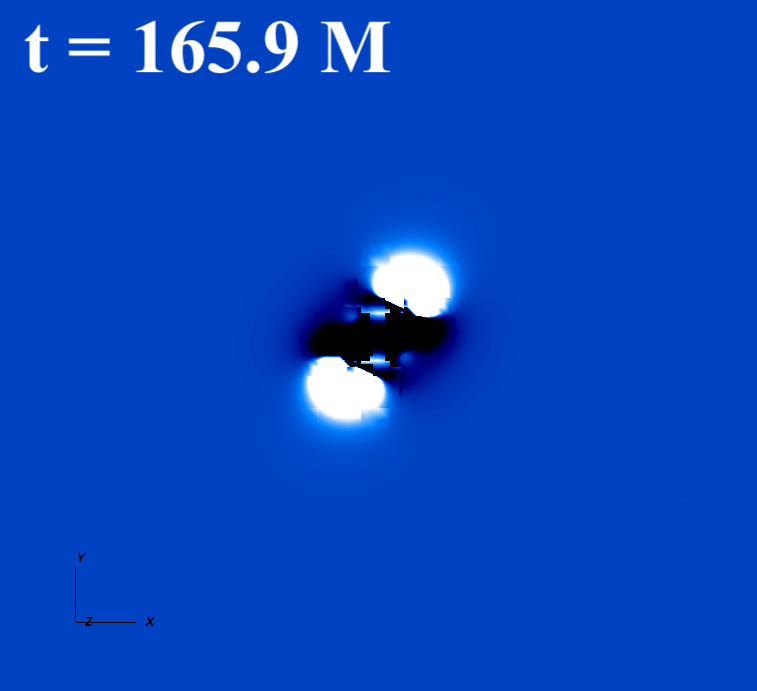}
\includegraphics[scale=0.093,trim=0.1cm 0cm 0cm 0cm]{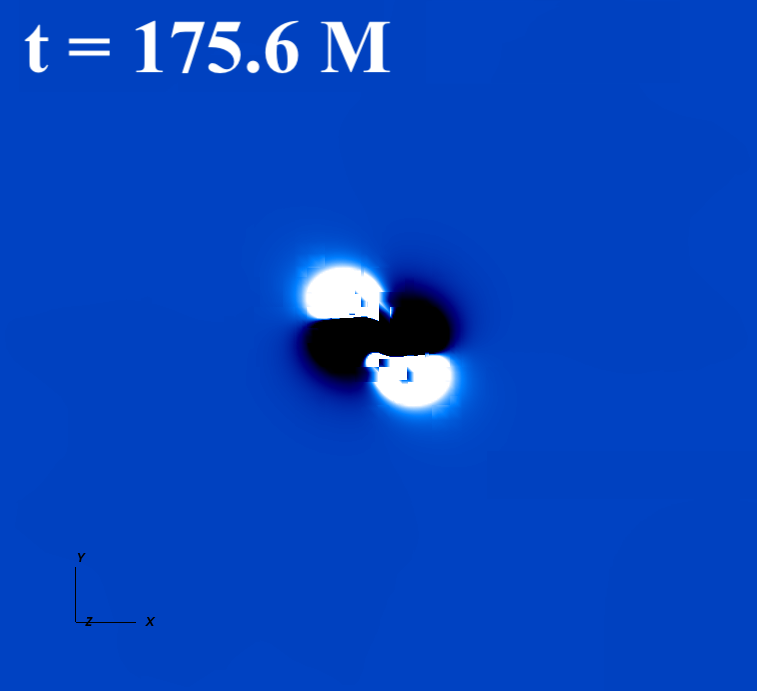}
\includegraphics[scale=0.093,trim=0.1cm 0cm 0cm 0cm]{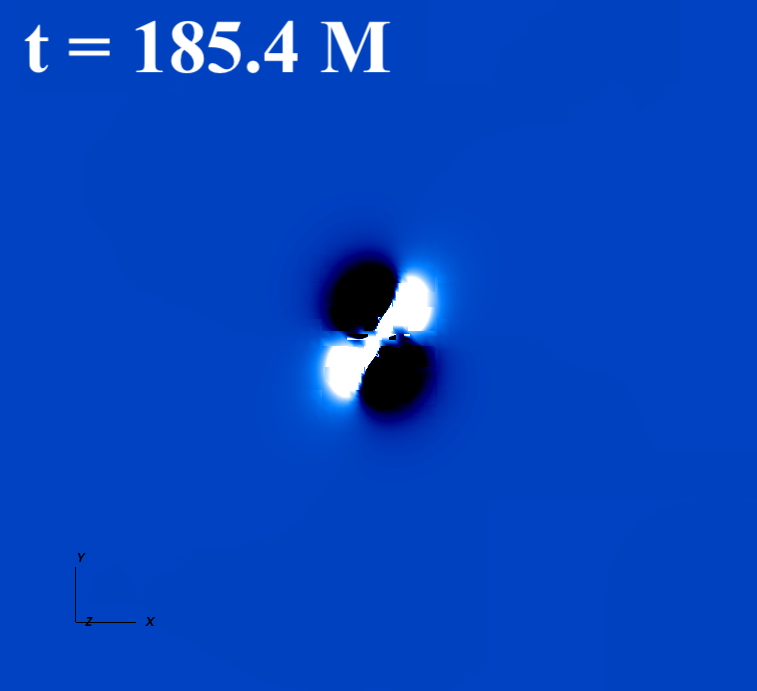}
\includegraphics[scale=0.093,trim=0.1cm 0cm 0cm 0cm]{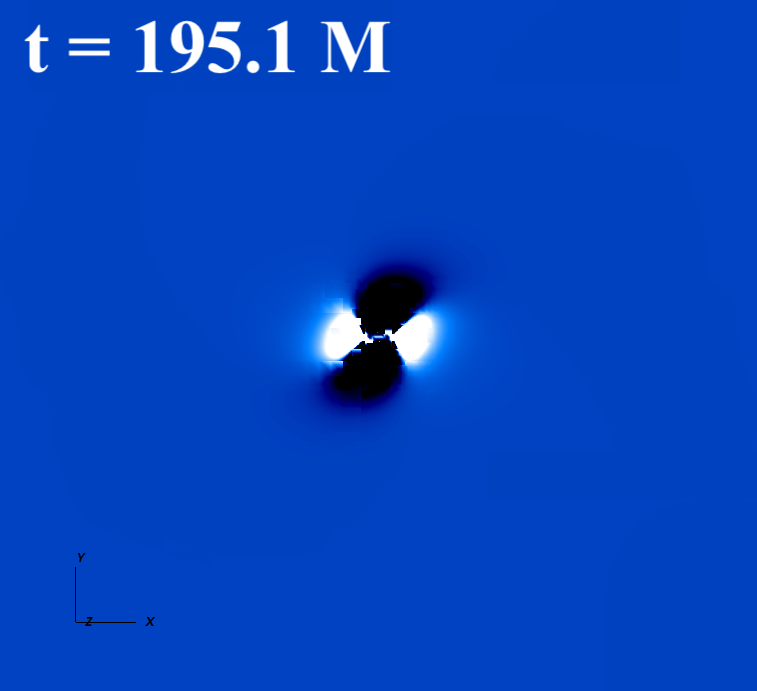}
\includegraphics[scale=0.093,trim=0.1cm 0cm 0cm 0cm]{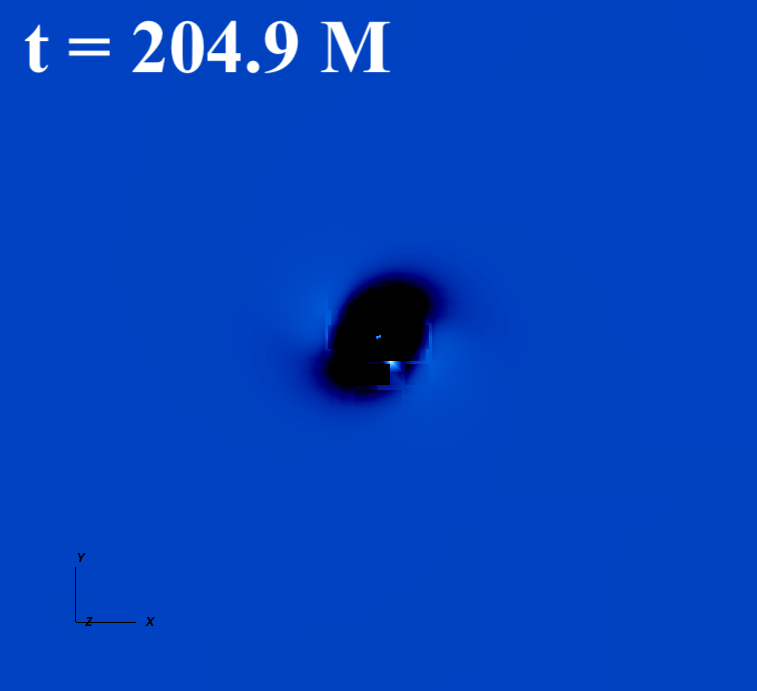}
\includegraphics[scale=0.093,trim=0.1cm 0cm 0cm 0cm]{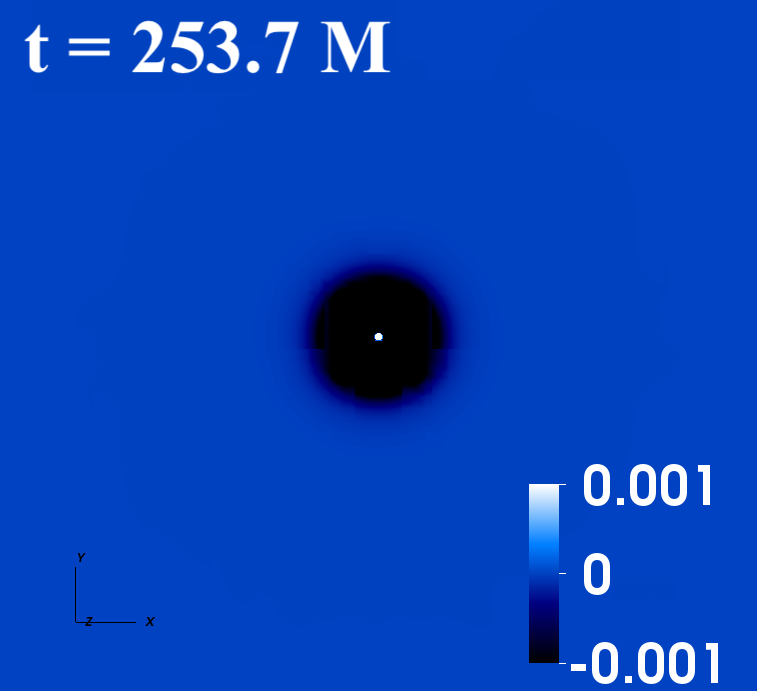}\\
\caption{Snapshots of the temporal evolution of  $E_{ab}B^{ab}$ in the $z=0$ coordinate plane for a BBH merger with parameters $q=1, D/2=8.1M$ and $a_{1,2}/m_{1,2}=0.63$. 
Time and length are expressed in units of the ADM mass $M$.}
\label{fig.3}
\end{figure}

\begin{figure}[h!]
\centering
 \includegraphics[scale=0.3,trim=0.1cm 0cm 0cm 0cm]{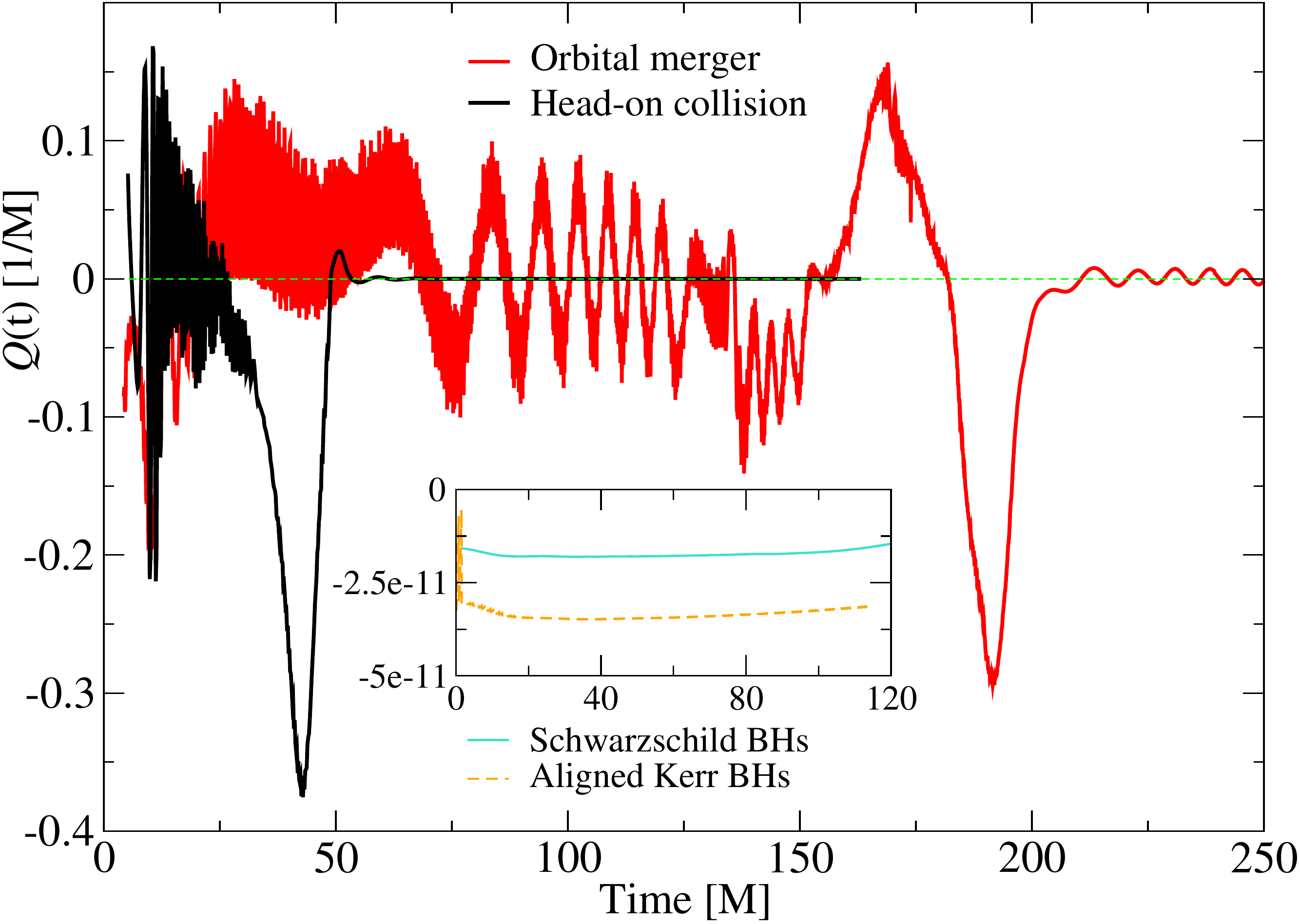}
\caption{Time evolution of the spatial integral in Eq.~(\ref{DQ}) for a representative orbital merger (red curve) and  head-on collision (black curve) of misaligned Kerr BHs. The inset corresponds to orbital mergers of both Schwarzschild  and aligned Kerr BHs,  compatible with zero for all time. 
 {Units are expressed in terms of  the ADM mass $M$.} }
\label{fig.4}
\end{figure}

To gain further intuition, Figure~\ref{fig.3} shows snapshots of  the evolution of the function $E_{ab}B^{ab}(x,y)$ from the inspiral-merger-ringdown phases of two Kerr BHs with the same spin magnitude and $D/2=8.1M$.  The time evolution of { {\it Q}$(t)$:=}$\int_{\Sigma_t} d^3x \sqrt{-g} E_{ab}B^{ab}$ is shown in the red curve of  Fig.~\ref{fig.4}.  
Using symmetry considerations as above we can qualitatively understand its behaviour  in the last two orbits of the inspiral. At $t\sim 138.2M$ the spins of the two BHs are pointing inwards the orbit (see Fig. \ref{fig.3}),  corresponding to a maximal configuration, thus explaining the negative peak found. At $\sim 158.5M$ the two objects have evolved one quarter of an orbit, and their spins have acquired a relative orientation that is even under mirror reflections with respect to the separation plane, leading to a zero value of (\ref{DQ}). When  the system has evolved another quarter of an orbit, at  time $\sim 166.7M$, the two spins acquire again maximal  configuration  but now pointing outwards the orbit;  this is the mirror-reflected initial configuration and explains the observed positive maximum peak.
After another quarter of an orbit (at $\sim 182.9M$) the two objects reach again a configuration that is even under mirror reflections, producing another zero value. Returning  to the initial configuration after  a full cycle (at $\sim 191.1M$), the spatial integral attains  the maximum negative peak observed, right {before the merger}. The amplitude of the oscillations grows because the relative distance between the BHs decreases significantly in the last orbits.  
After merger the value of (\ref{DQ}) decays to zero, as corresponds to a final Kerr BH.

 In the inset of Fig.~\ref{fig.4} we show the time evolution of $Q(t)$ for the orbital mergers of two Schwarzschild BHs and two Kerr BHs with spins aligned with the orbital plane. In both cases the result is zero up to roundoff error for any time, as expected from  symmetry considerations.
 This confirms the conclusions drawn in the previous section using waveform catalogues. 

Our simulations show that the main contribution to the Chern-Pontryagin in a BBH merger is attained around the time of merger.  
{The behaviour at merger} can be qualitatively understood by analyzing head-on collisions, which are computationally cheaper than orbital mergers,  allowing us to study the direct influence of the  BH spin values in a much reduced parameter space. For this reason, we consider  head-on collisions of Kerr BHs, initially separated by $D/2=8.3M$, and with anti-aligned spins in the $x$ direction, i.e.~$(\rightarrow,0,0),\, (\leftarrow,0,0)$.  For all models considered, a peak in the  evolution of $Q(t)$ is found around the time of collision, as shown in the black curve of Fig.~\ref{fig.4}. The total integrated effect is given in Table~\ref{tabhdbbh}, which also reports the GW contribution to the Chern-Pontryagin for comparison.

\begin{table}
\begin{center}
\begin{tabular}{|c|c|c|c|c|c|}
\hline
Type&$q$&$D/2M$&$a_{i}/m_{i}$  & $\Delta Q_{\mathcal J^+}$  &  $\left<Q_D(\mathcal J^+)  \right>$ \\
\hline
Orbital NS&1&2.5&0.00& 2.0$\times10^{-15}$ & 3.4$\times10^{-11}$\\
Orbital A&1&2.1&0.62& 1.2$\times10^{-14}$ & 6.4$\times10^{-11}$\\
Orbital P&1&4.5 &  0.63&    1.2$\times10^{-4}$   & 0.08       \\
Orbital P&1&8.1  &  0.63&    1.3$\times10^{-4}$  &  0.02     \\
Head-on &1&8.3  &  0.63&    5.1$\times10^{-6}$  &   0.07    \\
\hline

\end{tabular}
\end{center}
\caption{ Values of $\left<{\rm 0}|Q_D(\mathcal J^+)  |{\rm 0}\right>$ computed from (\ref{DQ})  with numerical-relativity simulations of representative orbital mergers and head-on collisions of  BHs. The  contribution from GW, $\Delta Q_{\mathcal J^+}$, is given  for comparison. 
\label{tabhdbbh}} 
\end{table}

{\bf {\em Discussion}}.
The close link between {chiral anomalies} and circular polarization of GW established in (\ref{CPintegral}) and (\ref{eb_eq})-(\ref{ebu}) implies that a chiral flux of GW induces the excitation from the quantum vacuum of a flux of photons  with {\it net circular polarization}, $\left<{\rm 0}|Q_D(\mathcal J^+)  |{\rm 0}\right>\neq0$. This is not accounted by  gravitational instantons, since no change of topology occurs. Furthermore, for any numerical simulation in relativistic astrophysics, (\ref{eb_eq})-(\ref{ebu}) can be used to identify the emergence of this quantum effect by simply relying on the resulting gravitational waveforms. {In fact, the chiral flux of photons can be evaluated  from LIGO/Virgo detections directly.} This work, hence, builds  a bridge between numerical relativity, GW observations, and the physical predictions of quantum field theory in curved spacetimes.

 Although the Hawking effect for  stationary BHs  predicts a helicity-dependent angular distribution \cite{Vilenkin-Unruh}, the net contribution when integrated over all angles is zero, in agreement with the vanishing of (\ref{CPintegral})  for Kerr metric. We report here a net creation of helicity during collisions of misaligned BHs, where all spacetime mirror symmetries are broken,  peaking at the time of merger  with value $5\times 10^{-3}/M$. This is comparable to the total  Hawking's emission rate of photons  by  the final BH configuration,  $10^{-3}/M$ \cite{Page76}. Note in passing that both effects, although intrinsically different, scale similarly. 

The results obtained 
for astrophysical BBH coalescences are  small to be observed directly, mainly due to the quantum-gravitational origin of the effect. However, its emergence  can act as a ``seed'' that, together with possible classical mechanisms of amplification, might lead to  more observable predictions. Without  such mechanisms, the amplitude of this quantum effect seems of observational relevance only if there exists an accumulative process over long periods of time, {or a large number of observed events, as expected in the near future from LIGO/Virgo interferometers.} 
One example could be the  inspiral phase of compact binaries, which last millions of years. The contribution from BBH mergers with adiabatic orbital precession is expected to be negligible because the spin-configuration at different times along the orbit are related by a mirror reflection, and consequently they cancel each other when integrating in time; only highly (orbital and/or spin) precessing inspirals could be important. 
On the other hand, BNS systems, that have a richer multipolar structure and can be endowed with strong magnetic fields, can avoid the cancellations in time and  lead to predictions of interest. Let us suppose an idealized  BNS system in which each star has a misaligned rotation axis and magnetic field axis. If the rotation of the components is synchronized with the orbital motion, the magnetic vector field of each star will spin around during the orbital motion at the same rate. The spatial configuration of the four axial vectors will thus remain approximately invariant during the evolution. 
Eqs.~(\ref{eb_eq})-(\ref{ebu})   yield an estimate of $\Delta Q_{\mathcal J^+}\approx 10^3$ in over 1-2 years of observation, by assuming that the contribution of (\ref{ebu}) is in the same ballpark as for the precessing BBH case of Fig.~\ref{fig.2}.
Future developments of BNS numerical simulations  involving higher (and misaligned) spin values and/or magnetic fields will help to quantify this effect in realistic scenarios and to estimate its chances of being measured in multimessenger astronomy observations of BNS mergers such as GW170817~\cite{GW170817_1}

 Our analysis  finds applications in other contexts as well. For instance, 
  the physical effect described in this paper can be straightforwardly extended to fermions, since they also have a chiral anomaly dictated by the Chern-Pontryagin \cite{kimura69}. Therefore, a flux of circularly polarized GW (or electromagnetic waves \cite{ABJ, footnote-4}) can also trigger the creation of chiral fermions. For neutrinos this is equivalent to a creation of leptonic asymmetry.
   Thus, the coalescence of primordial BH in the early universe, when lepton and baryons can be interconverted, could  play a role   in the matter-antimatter asymmetry without requiring  any  assumption beyond standard quantum field theory and general relativity \cite{APSJ06}. On the other hand, our results could also be applied to analyze the stability of magnetic fields in neutron stars \cite{Spruit08}.  A non-vanishing magnetic helicity is needed to avoid the decay of magnetic fields, but its origin  remains to be understood.
     Using  \cite{DKZ87} it can be  deduced that a non-zero Chern-Pontryagin produces a non-zero magnetic helicity by quantum fluctuations, and according to our results this may be  achieved  in a core colapse supernovae emitting highly polarized GW.

At a more speculative level,  there are some astrophysical systems that emit circularly polarized light in which this quantum effect might also be operating. 
Some fast radio bursts (FRBs), a class of very bright and short (milliseconds) electromagnetic bursts, whose origin remains unknown~\cite{Katz}, show significant intrinsic circular polarization \cite{Petroff1,Petroff3}. Our results show that for stellar-mass BBH mergers there is a peak in the V-Stokes parameter that lasts for a few miliseconds, thus resembling qualitatively  the observed polarization profiles in FRBs  (compare our Fig.~3 with Fig.~12 of~\cite{Petroff3}). Given that it is plausible that BH play a crucial role in the physics of  FRBs, it would be interesting to investigate whether the mechanics studied in this paper is related to these observed phenomena.

As a byproduct of this work, we notice that circular polarization of GWs can be a  useful measurement from  LIGO/Virgo observations to extract information about BBH mergers since, as shown above,  
its detection  would be a clear indication on the existence of precession, something that usually has only a weak imprint on the observable signal \cite{Fairhurstetal19}.

{\bf Acknowledgements.} We are grateful to T. Dietrich, V. Chaurasia and collaborators for providing waveform data from their BNS simulations. We thank A. Ashtekar, E. Bianchi, M. Campanelli, M. Campiglia, V. Cardoso, M. Casals, F. Duque, G. Faye, S. Gimeno-Soler, C. Herdeiro, A. Torres-Forne,  J. Zanelli, and Y. Zlochower for useful comments. This work also benefited from discussions  during the GR22/Amaldi13 conference in Valencia. 
ADR  acknowledges financial support by the European Union's ERC Consolidator Grant ``Matter and strong-field gravity: New frontiers in Einstein's theory'' no. MaGRaTh--646597; NSG by the Funda\c c\~ao para a Ci\^encia e a Tecnologia (FCT) project PTDC/FIS-OUT/28407/2017; JAF by the Spanish Agencia Estatal de Investigaci\'on  (grant PGC2018-095984-B-I00), the  Generalitat  Valenciana  (PROMETEO/2019/071) and the European Union’s Horizon 2020 RISE programme (H2020-MSCA-RISE-2017  Grant  No.~FunFiCO-777740); IA by the NSF CAREER grant PHY- 1552603, and   the Hearne Institute for Theoretical Physics of Louisiana State University; VM by NSF Grants No.\  OAC-1550436, AST-1516150, PHY-1607520, PHY-1305730, PHY-1707946,  PHY-1726215, the Exascale Computing Project (17-SC-20-SC), funds from AYA2015-66899-C2-1-P, and RIT for the FGWA SIRA initiative; and JNS by the Ministerio de Economia y Competitividad grants no FIS2017-91161-EXP and FIS2017-84440-C2-1-P.


\begin{thebibliography}{99}
\bibitem{footnote-1} The actual  number operator at $\mathcal J^+$ representing the difference between right- and left-handed photons is given by the ``normal-ordered'' operator $Q_D(\mathcal J^+)- \mathbb I \left<{\rm out}|Q_D(\mathcal J^+)|{\rm out}\right>_{\rm ren}$, where $\left|{\rm out}\right>$ denotes the preferred vacuum state at  $\mathcal J^+$. But  $\left<{\rm out} |Q_D(\mathcal J^+)|{\rm out} \right>_{\rm ren}=0$.
\bibitem{footnote-2.5} The  choice $\omega'=0$ recovers the well-known Chern-Simons 3-form for $T$.  We keep $\omega'$ general, since we shall make use of a  convenient class of coordinate systems used in \cite{NU68} (leading to $\omega'\neq 0$)  that makes relatively simple the asymptotic behaviour of the spin-coefficients.
\bibitem{footnote-3}
By taking this auxiliary Minkowski metric we are implicitly assuming that our manifold is $M\simeq U\subset \mathbb R^4$ (open $U$, with the usual topology), since other  $M$ may not admit a flat connection globally. The topology in the problem is thus trivial, and this is supported by the topological censorship theorem in general relativity \cite{FSW93}.
\bibitem{footnote-4}  To gain  intuition about this expression, one can work out  the electromagnetic analogue first. In that case, the Chern-Pontryagin integral is given by $\int_M F_{ab}{^*F}^{ab} d^4x=\int_{\mathcal J^+} du \, d\mathbb S^2  \int^{u'} du' [ {\rm Re}  \phi_2^0(u) {\rm Im}\bar\phi_2^0(u') -  {\rm Re}  \phi_2^0(u') {\rm Im}\bar\phi_2^0(u)]$, where $\phi_2$ is the analogue of $\Psi_4$ in the Maxwell case. This result represents the difference between right- and left-handed photons crossing future null infinity once the fields are expanded in Fourier modes.
\end{thebibliography}
\end{document}